\documentclass[aps,prx,twocolumn,floatfix,longbibliography,superscriptaddress]{revtex4-2}

\usepackage{amsmath}
\usepackage{amssymb}
\usepackage{times}
\usepackage{braket}
\usepackage[pdftex]{graphicx}
\usepackage{color}
\usepackage{blindtext}
\usepackage{nicefrac}
\usepackage[colorlinks=true, citecolor=blue, urlcolor=blue, linkcolor=blue]{hyperref}

\renewcommand{\vec}[1]{\mathbf{#1}}

\newcommand{\change}[1]{\textcolor{black}{#1}}
\newcommand{\changeb}[1]{\textcolor{black}{#1}}

\renewcommand{\ket}[1]{\lvert#1\rangle} 
\newcommand{\braopket}[3]{\langle #1 | #2 | #3\rangle} 

\begin{document}

\title{Topological orbital Hall effect caused by skyrmions and antiferromagnetic skyrmions}

\author{B{\"o}rge G{\"o}bel}
\email[Correspondence email address: ]{boerge.goebel@physik.uni-halle.de}
\affiliation{Institut f\"ur Physik, Martin-Luther-Universit\"at Halle-Wittenberg, D-06099 Halle (Saale), Germany}

\author{Lennart Schimpf}
\affiliation{Institut f\"ur Physik, Martin-Luther-Universit\"at Halle-Wittenberg, D-06099 Halle (Saale), Germany}

\author{Ingrid Mertig}
\affiliation{Institut f\"ur Physik, Martin-Luther-Universit\"at Halle-Wittenberg, D-06099 Halle (Saale), Germany}

\date{\today}

\begin{abstract}
\noindent \textbf{Abstract.}
The topological Hall effect is a hallmark of topologically non-trivial magnetic textures such as magnetic skyrmions. It quantifies the transverse electric current that is generated once an electric field is applied and occurs as a consequence of the emergent magnetic field of the skyrmion. Likewise, an orbital magnetization is generated. Here we show that the charge currents are orbital polarized even though the conduction electrons couple to the skyrmion texture via their spin. The topological Hall effect is accompanied by a topological orbital Hall effect even for $s$ electrons without spin-orbit coupling. As we show, antiferromagnetic skyrmions and antiferromagnetic bimerons that have a compensated emergent field, exhibit a topological orbital Hall conductivity that is not accompanied by charge transport and can be orders of magnitude larger than the topological spin Hall conductivity. \changeb{Skyrmionic textures serve as generators of orbital currents that can transport information and give rise to considerable orbital torques.}
\end{abstract}

\maketitle


\noindent\textbf{Introduction}\\
Magnetic skyrmions are non-collinear spin textures that possess an innate stability due to their non-trivial real-space topology~\cite{bogdanov1989thermodynamically,muhlbauer2009skyrmion,nagaosa2013topological}. They have been observed as individual magnetic objects~\cite{romming2013writing,sampaio2013nucleation,fert2013skyrmions,yu2010real} that may serve as carriers of information in future storage technologies~\cite{sampaio2013nucleation,fert2013skyrmions}. At finite temperatures and magnetic fields, they even form periodic lattices~\cite{yu2010real,heinze2011spontaneous} with a rather homogeneous topological charge density $n_\mathrm{Sk}$.

This topological charge density has been identified with an effective magnetic field $\vec{B}_\mathrm{em}\propto n_\mathrm{Sk}\vec{e}_z$, called `emergent field'~\cite{schulz2012emergent,nagaosa2013topological} that affects the conduction electrons by changing the phase of the wave function; a Berry phase is accumulated~\cite{berry1984quantal,zak1989berry}. While the electron spins align with the skyrmion texture, an effective Lorentz force acts on their charges and leads to the emergence of transverse transport phenomena: The topological Hall effect~\cite{bruno2004topological,neubauer2009topological,lee2009unusual} describes the emergence of a transverse charge current once an electric field is applied. It is the hallmark of the skyrmion crystal phase and the magnitude of its resistivity can be used to measure the skyrmion and topological charge densit\changeb{ies}~\cite{maccariello2018electrical,sivakumar2020topological,raju2021colossal}. Since the spins align with the skyrmion texture, the currents are also spin polarized and a topological spin Hall conductivity has been predicted~\cite{yin2015topological,ndiaye2017topological,gobel2018family}.
When two skyrmions with opposite topological charges are coupled to form an antiferromagnetic skyrmion~\cite{barker2016static,zhang2016magnetic,zhang2016antiferromagnetic,gobel2017afmskx,legrand2020room,dohi2019formation}, the topological spin Hall effect is still present but it is not anymore accompanied by charge transport~\cite{buhl2017topological,gobel2017afmskx}.

Different types of skyrmions have been observed on several different length scales~\cite{gobel2021beyond} and even nano-sized skyrmions can be stabilized~\cite{heinze2011spontaneous,okubo2012multiple}. In this case, the emergent field has a magnitude of several thousand Tesla. As a consequence, quantized transport effects occur~\cite{hamamoto2015quantized,gobel2017THEskyrmion,gobel2017QHE} akin to the quantum Hall effect in the presence of a large magnetic field~\cite{landau1930diamagnetismus,onsager1952interpretation,hofstadter1976energy,klitzing1980new,thouless1982quantized,hatsugai2006topological,sheng2006quantum}. In fact, if the coupling between conduction electron spins and the skyrmion texture is strong, the system can be mapped to a quantum Hall system~\cite{hamamoto2015quantized,gobel2017THEskyrmion,gobel2017QHE} and Landau levels occur giving rise to edge states in the form of skipping orbits.

Over the recent years, the orbital degree of freedom has become increasingly relevant in the field of spinorbitronics, manifesting itself in phenomena such as orbital magnetization~\cite{chang1996berry,xiao2005berry,thonhauser2005orbital,ceresoli2006orbital,raoux2015orbital,gobel2018magnetoelectric}, orbital torque~\cite{go2020orbital,lee2021orbital}, orbital Edelstein effect~\cite{johansson2021spin,el2023observation} and orbital Hall effect~\cite{zhang2005intrinsic, bernevig2005orbitronics, kontani2008giant, tanaka2008intrinsic, kontani2009giant,go2018intrinsic, go2021orbitronics,pezo2022orbital,canonico2020orbital,cysne2022orbital,salemi2022theory,busch2023orbital,choi2023observation,cysne2021disentangling,barbosa2023orbital,seifert2023time,busch2024ultrafast,gobel2024OHE, sahu2021effect, bhowal2021orbital, salemi2022first, sala2022giant, pezo2023orbital, go2024first, wang2024orbital, liu2024quantum, atencia2024orbital, liu2024dominance}. In the presence of a magnetic field, charge currents form circular trajectories and generate an orbital angular momentum that can be calculated via the modern formulation of orbital magnetization~\cite{chang1996berry,xiao2005berry,thonhauser2005orbital,ceresoli2006orbital,raoux2015orbital,gobel2018magnetoelectric}. This orbital magnetization has been considered for quantum Hall systems as well as skyrmions~\cite{dos2016chirality,lux2017chiral,gobel2018magnetoelectric}. Recently, we have shown that the skipping orbits in quantum Hall systems lead to orbital-polarized edge currents~\cite{gobel2024OHE} that give rise to an orbital Hall effect accompanying the (charge) Hall effect once a transverse electric field is applied. 


\begin{figure}[t!]
    \centering
    \includegraphics[width=0.8\columnwidth]{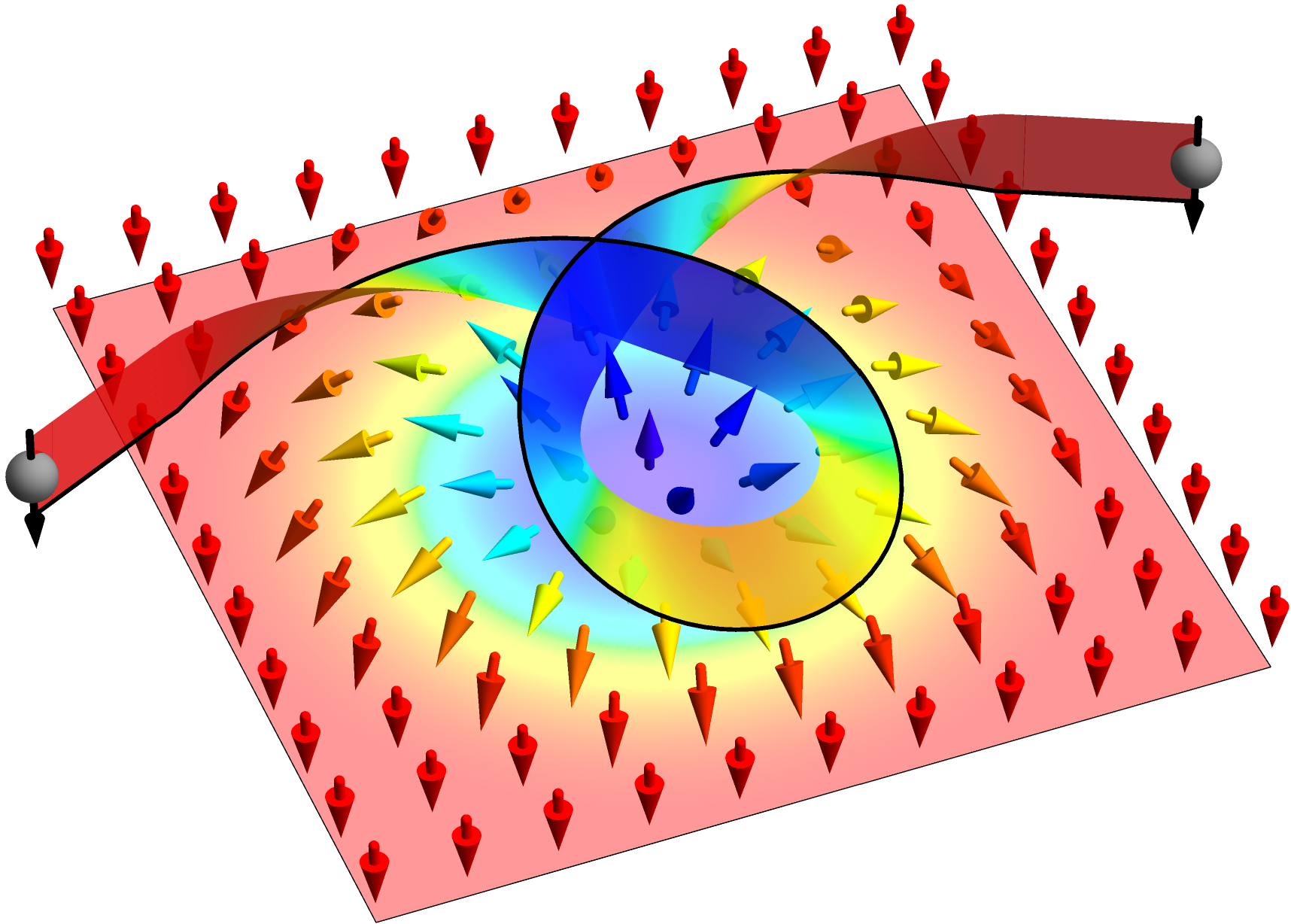}
    \caption{\textbf{Topological orbital Hall effect.} The spin of the conduction electron (black) aligns with the skyrmion texture (colored arrows; the color resembles the out-of-plane orientation). While moving through the skyrmion, the electron accumulates a Berry phase and is deflected (topological Hall effect). Due to the cycloid trajectory (black), an orbital angular momentum is generated and transported as an orbital current (topological orbital Hall effect).}
    \label{fig:overview}
\end{figure}

\begin{figure*}[t!]
    \centering
    \includegraphics[width=\textwidth]{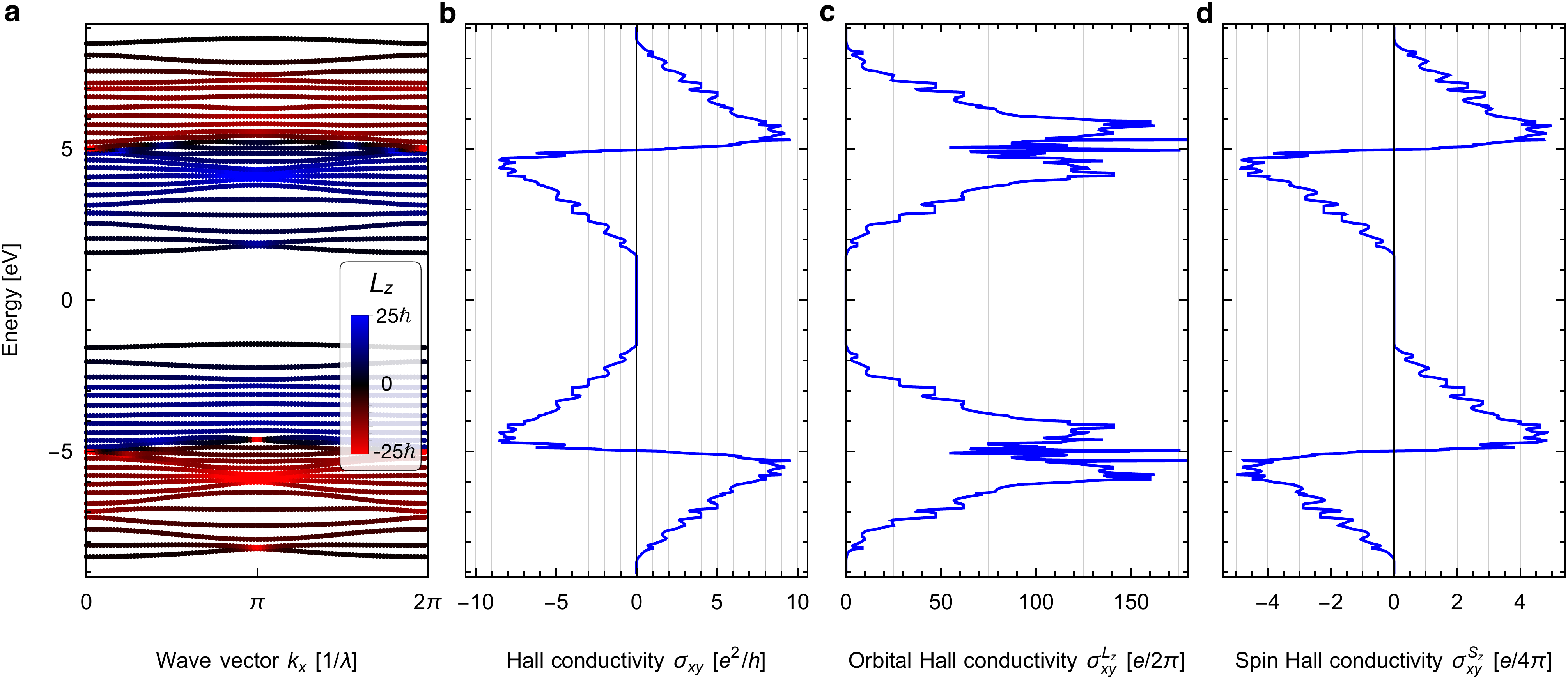}
    \caption{\textbf{Electronic transport in skyrmion crystals.} \textbf{a} Band structure $E_{\nu\vec{k}}$. The color code indicates the $\vec{k}$- and band-resolved out-of-plane orbital angular momentum $L_{z,\nu}(\vec{k})$ (blue: positive, red: negative). The band structure consists of 2 blocks corresponding to parallel and anti-parallel spin alignment. \textbf{b} Hall conductivity $\sigma_{xy}$ as a function of energy. It is quantized in units of $e^2/h$ in the band gaps. \textbf{c} Orbital Hall conductivity $\sigma^{L_z}_{xy}$ as a function of energy. \textbf{d} Spin Hall conductivity $\sigma^{S_z}_{xy}$ as a function of energy. These calculations correspond to a skyrmion size of $\lambda=5a$, and a Hund's coupling of $m=5|t|=5\,\mathrm{eV}$.}
    \label{fig:m5bulk}
\end{figure*}

In this paper, we show that skyrmions give rise to a topological orbital Hall effect (Fig.~\ref{fig:overview}). The orbital currents appear additionally to the topological (charge) Hall effect giving rise to transverse orbital-polarized currents. By using a tight-binding model and a Berry curvature approach, we systematically compare the charge, spin and orbital Hall conductivities. Furthermore, we analyze the edge states that occur when the skyrmion is on the nanometer scale. In this scenario the emergent field of the skyrmion is extremely large and Landau levels form. 
As we show, when two oppositely magnetized skyrmions are coupled to form an antiferromagnetic skyrmion, the topological Hall effect is compensated and only orbital and spin Hall effects emerge. Since the orbital angular momentum of the conduction electrons in a skyrmion system \changeb{can be arbitrarily high}, the topological orbital Hall conductivity can be orders of magnitude larger than the topological spin Hall conductivity.\\
%
%
%
%
\\
\noindent\textbf{Results and discussion}\\
\noindent\textbf{Spinorbitronic transport in non-collinear spin textures.}
We consider a square lattice with lattice constant \changeb{$a=2.76\,$\AA} and a single $s$ orbital per site for the conduction electrons. The skyrmion texture is a rotational symmetric vector field $\vec{m}(\vec{r})$ (normalized) with diameter $\lambda$ which is shown in Fig.~\ref{fig:overview} (for details see Methods section) that is formed by energetically lower lying $d$ electrons. The non-collinear texture increases the magnetic unit cell to $(\lambda/a)\times(\lambda/a)$ lattice sites. 

The $s$-$d$ Hamiltonian consists of a nearest-neighbor hopping term (amplitude $t=-1\,\mathrm{eV}$) and a Hund's coupling term (quantified by $m$)
\begin{align} 
  H &  =  \sum_{\braket{ij}} t \,c_{i}^\dagger c_{j} + m \sum_{i} \vec{m}_{i} \cdot (c_{i}^\dagger \boldsymbol{\sigma}c_{i}).
  \label{eq:ham_the} 
\end{align} 
$c_{i}^\dagger$ and $c_{i}$ are the creation and annihilation operators at site $i$. $\boldsymbol{\sigma}$ is the vector of the Pauli matrices characterizing the spin of the conduction electrons. The eigenvalues $E_{\nu\vec{k}}$ are the band structure (band index $\nu$ and wave vector $\vec{k}$) and the eigenvectors $\ket{\nu \vec{k}}$ will be used to determine the observables discussed in this paper: Spin $S_{\nu,z}(\vec{k})$ and orbital angular momentum $L_{\nu,z}(\vec{k})$, as well as the charge Hall conductivity $\sigma_{xy}(E_\mathrm{F})$, spin Hall conductivity $\sigma_{xy}^{S_z}(E_\mathrm{F})$ and orbital Hall conductivity $\sigma_{xy}^{L_z}(E_\mathrm{F})$ that are calculated as integrals over the Berry curvature $\Omega_{\nu,z}(\vec{k})$, spin Berry curvature $\Omega_{\nu,z}^{S_z}(\vec{k})$ and orbital Berry curvature $\Omega_{\nu,z}^{L_z}(\vec{k})$ over the Brillouin zone as functions of the Fermi energy $E_\mathrm{F}$. For more details about the calculations, we refer to the Methods section. 

\begin{figure*}[t!]
    \centering
    \includegraphics[width=\textwidth]{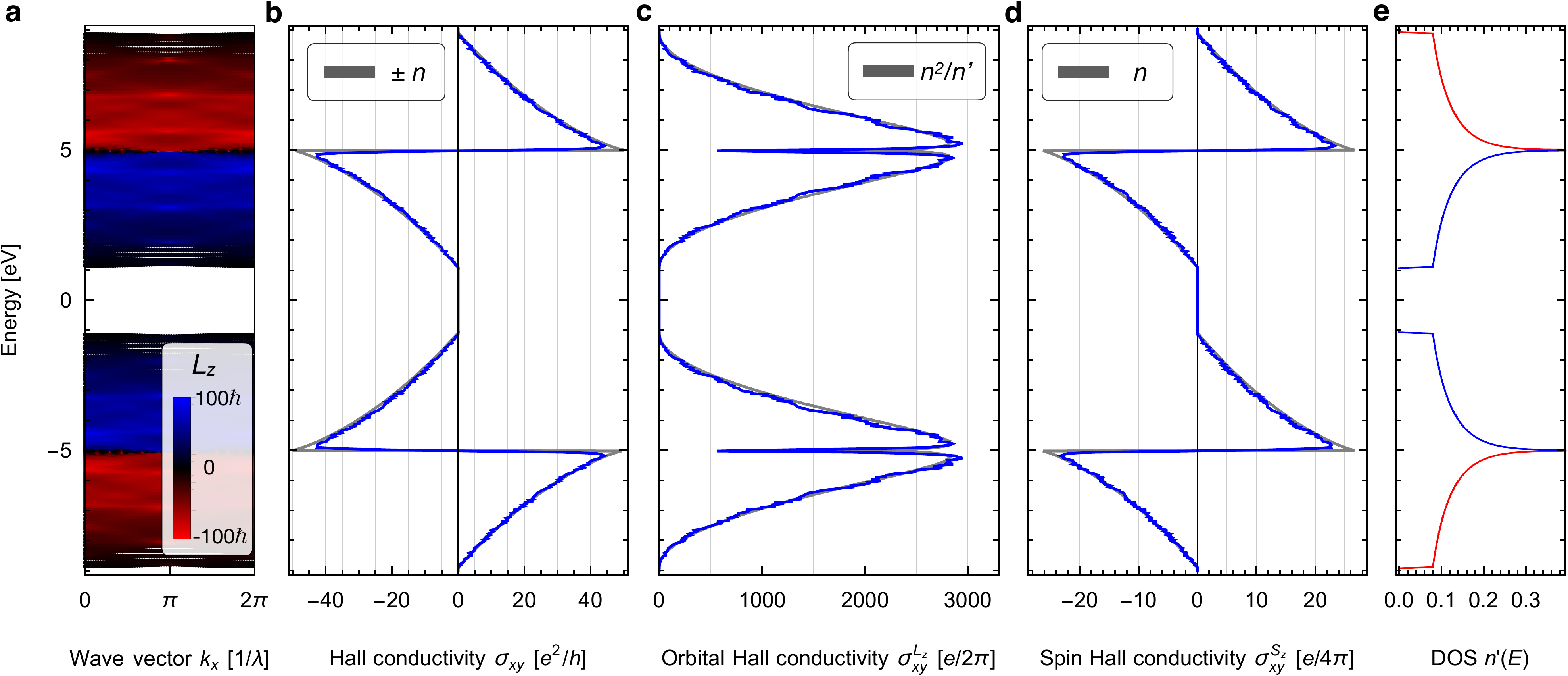}
    \caption{\change{\textbf{Simplified explanation of transport properties.} \textbf{a} Band structure $E_{\nu\vec{k}}$. The color code indicates the $\vec{k}$- and band-resolved out-of-plane orbital angular momentum $L_{z,\nu}(\vec{k})$ (blue: positive, red: negative). \textbf{b} Hall conductivity $\sigma_{xy}$ as a function of energy. \textbf{c} Orbital Hall conductivity $\sigma^{L_z}_{xy}$ as a function of energy. \textbf{d} Spin Hall conductivity $\sigma^{S_z}_{xy}$ as a function of energy. \textbf{e} Density of states $n'(E)$ of the band structure of the underlying lattice $E=2\tilde{t}[\cos(k_xa)+\cos(k_ya)]$ shifted by $\pm m$. The gray curves in b-d (scaled) resemble the trend expected based on this zero-field band structure: $\pm n(E)$ in panel b, $[n(E)]^2/n'(E)$ in panel c, and $n(E)$ in panel d. These calculations correspond to a skyrmion size of $\lambda=10a$ and a Hund's coupling of $m=5|t|=5\,\mathrm{eV}$.}}
    \label{fig:m5bulklarge}
\end{figure*}

\textbf{Band structure, orbital and spin angular momentum in skyrmion textures.}
The band structure $E_{\nu\vec{k}}$ depends on the strength of the Hund's coupling $m$. Starting the discussion from $m=0$, all bands are spin degenerate and the band structure resembles the single tight-binding band of the square lattice $E(\vec{k})=2t[\cos(k_xa)+\cos(k_ya)]$ but backfolded into the smaller magnetic Brillouin zone accounting for the skyrmion. Upon increasing $m$, the electron spins start to align with the skyrmion texture and the initially spin-degenerate bands start to split. The band structure for $m=1\,\mathrm{eV}=|t|$ exhibits many (avoided) band crossings and is shown in Fig.~S1 of the Supplementary Material.

Once the coupling $m$ is larger than the initial band width $4|t|$, the spins are almost completely aligned with the skyrmion. Two blocks emerge -- one for parallel spin alignment and one for antiparallel spin alignment -- shifted by $\pm m$, respectively. Results for $m=5\,\mathrm{eV}=5|t|$ are shown in Fig.~\ref{fig:m5bulk}(a). The bands in the two blocks are similar and have a weak dispersion. In the limit of strong Hund's coupling and large skyrmion sizes, the two blocks become completely equivalent. In this adiabatic limit, the bands are Landau levels akin to the bands of a quantum Hall system. In Refs.~\cite{gobel2017THEskyrmion,gobel2017QHE,gobel2018family}, we have shown that the Hamiltonian hosting the non-collinear skyrmion texture can be transformed to a quantum Hall Hamiltonian with a collinear magnetic field coupling to the charge of the electrons. This magnetic field is the `emergent field' $\vec{B}_\mathrm{em}$ mentioned in the introduction.

While the two blocks in the band structure are characterized by opposite spin alignment, that is rather homogeneous within a block ($S_{\nu,z}(\vec{k})\approx \pm \hbar/2\,\braket{m_z}$ with $\braket{m_z}$ the average normalized out-of-plane magnetic moment of the skyrmion), the orbital angular momentum $L_{\nu,z}(\vec{k})$ is also roughly opposite comparing the two blocks but changes within a block. We have added it as a color code to Fig.~\ref{fig:m5bulk}(a). Starting from the lowest band, it is negative and increases in magnitude until it changes sign near $E=-m=-5\,\mathrm{eV}$ and decreases back to zero. Its magnitude is much larger than the spin by multiple $\hbar$ (cf. legend in Fig.~\ref{fig:m5bulk}a).

\begin{figure*}[t!]
    \centering
    \includegraphics[width=\textwidth]{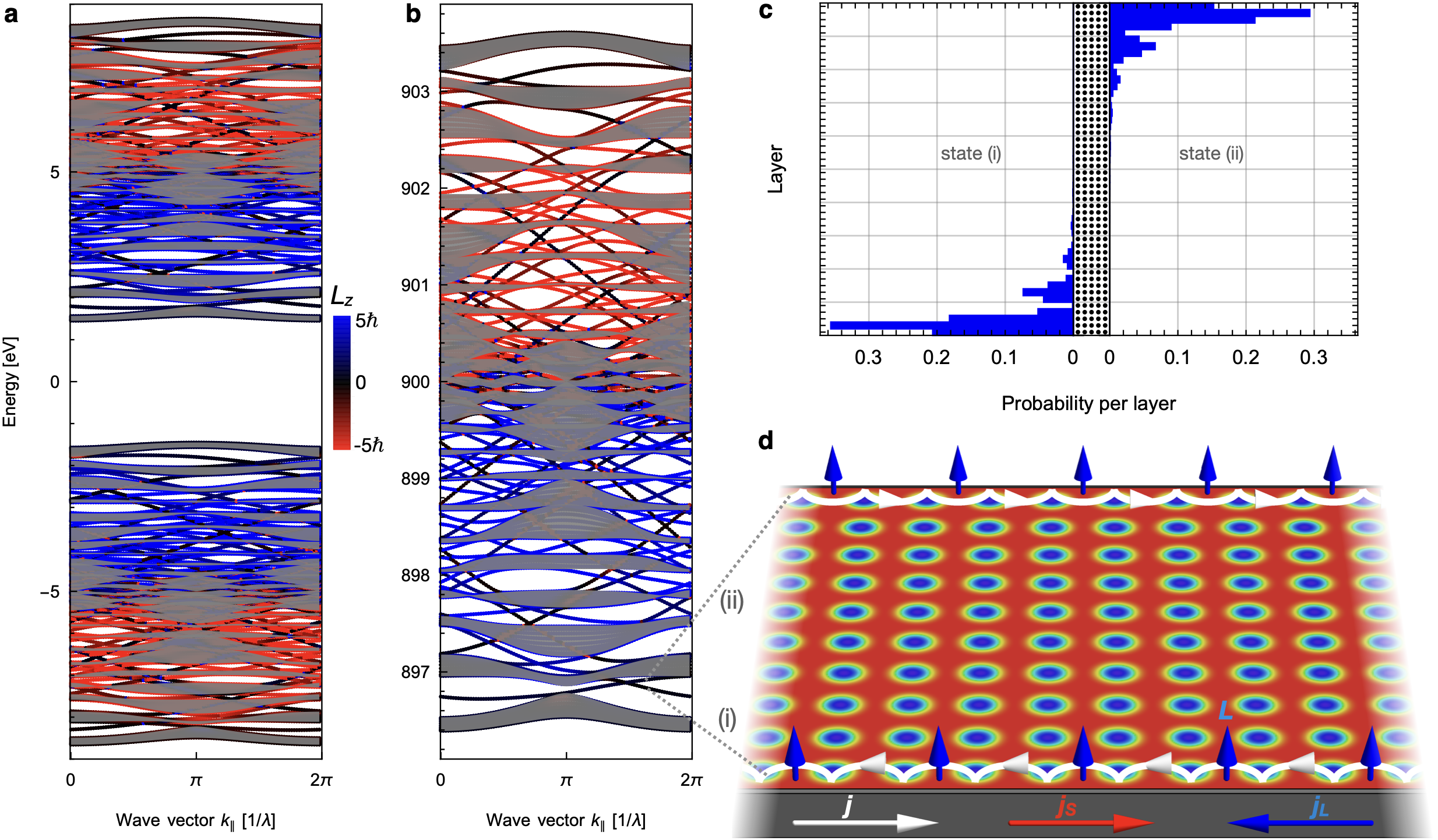}
    \caption{\textbf{Edge states carrying orbital angular momentum.} \textbf{a} Band structure of a slab with 10 skyrmions in the unit cell. The color indicates the orbital angular momentum (see legend). Gray areas indicate bulk states, projected from the bulk band structure (cf. Fig.~\ref{fig:m5bulk}a). These calculations correspond to a skyrmion size of $\lambda=5a$, and a Hund's coupling of $m=5|t|=5\,\mathrm{eV}$. \textbf{b} Zoom of the upper block for $m=900|t|=900\,\mathrm{eV}$ corresponding to the adiabatic limit of very strong coupling. \textbf{c} Layer-resolved probability of the two edge states highlighted in b, indicated as (i) and (ii) . The unit cell is indicated in the middle. \textbf{d} Schematic visualization of the two edge states giving rise to charge currents $\vec{j}$, spin currents $\vec{j}_S$ and orbital currents $\vec{j}_L$ that are oriented oppositely at both edges. \changeb{The magnetic skyrmions are resembled by the same color code as in Fig.~\ref{fig:overview}.}}
    \label{fig:edge}
\end{figure*}

\textbf{Topological charge, spin and orbital Hall effect.}
The analogy of the considered skyrmion system and a quantum Hall system allows to understand the emergence of a topological Hall effect, as discussed in Refs.~\cite{gobel2017THEskyrmion,gobel2017QHE}. Since the bands are (at least partially) spin polarized, the corresponding currents are spin polarized, as discussed in Ref.~\cite{gobel2018family}. Therefore, a topological spin Hall conductivity occurs as well. As we have just shown, the bands are also orbital polarized, so the emergence of a topological orbital Hall conductivity is expected as well. The energy dependent curves of the Hall conductivity $\sigma_{xy}$, orbital Hall conductivity $\sigma_{xy}^{L_z}$ and spin Hall conductivity $\sigma_{xy}^{S_z}$ are shown in Fig.~\ref{fig:m5bulk}(b-d), respectively. \change{Note that the following discussion is only true in the strong-coupling limit of the conduction electron spins and the magnetic texture. The results for a weaker coupling of $m=|t|=1\,\mathrm{eV}$ are shown in Fig.~S1 of the Supplementary Material.}

First, and most importantly, the orbital Hall conductivity (panel c) is non-zero. This means, a skyrmion exhibits orbital- and spin-polarized currents; a topological orbital Hall effect is superimposed on top of the previously observed topological Hall and topological spin Hall effects.

Both orbital (c) and spin Hall conductivities (d) can exhibit plateaus, e.\,g. between the two Landau levels near $E=3\,\mathrm{eV}$. However, the values are not quantized by a natural constant, because spin and orbital angular momentum are no good quantum numbers in the skyrmion system. The topological Hall conductivity (b) on the other hand is quantized in units of $e^2/h$ due to the mathematical equivalence to a quantum Hall system.

Since the electron spins align parallel with the skyrmion in the lower block of the band structure and antiparallel in the upper block, we see also two \change{b}locks in the conductivities. For the topological Hall effect (b), the signals are roughly reversed, because the parallel and antiparallel alignment corresponds to an interaction with oppositely oriented emergent fields $\vec{B}_\mathrm{em}\parallel\pm\vec{e}_z$ of the skyrmion. \changeb{Both} orbital (c) and spin Hall conductivities (d) exhibit almost the same signals comparing the two blocks because reversing the alignment of the spins with the skyrmion texture does not only reverse the emergent field but also $S_z$ and $L_z$. 

Within a \change{b}lock, the topological Hall conductivity (b) and spin Hall conductivity (d) change sign. This is because the electrons in the lower half of a block are characterized by a positive effective mass and in the upper half by a negative effective mass (hole-like behavior) which is determined by the band structure of the underlying square lattice. Particles of opposite mass are deflected into opposite directions by the same emergent magnetic field. However, by changing the sign of the effective mass, $L_z$ changes sign as well, as in classical physics. For this reason, the orbital Hall conductivity (c) remains always positive.

Fig.~\ref{fig:m5bulklarge} shows results analogous to Fig.~\ref{fig:m5bulk} but for an increased skyrmion diameter of $\lambda=10a$. The main difference is that there are more Landau levels and that the topological, orbital and spin Hall conductivities as well as the orbital angular momentum are increased compared to $\lambda=5a$. The scaling of the Hall conductivities with the skyrmion area is shown in Fig.~S2 in the Supplementary Material. While the topological Hall and spin Hall conductivities increase linearly with the skyrmion area, the orbital Hall conductivity increases quadratically.

\change{The trend of the energy dependencies of the Hall conductivities can even be quantified based on the analysis of the zero-field band structure of the underlying square lattice $E(\vec{k})=2t[\cos(k_xa)+\cos(k_ya)]$ shifted by $\pm m$. We relate the Hall conductivities to the carrier density $n(E)=\int_{-\infty}^E n'(E')\,\mathrm{d}E'$ and the density of states $n'(E)$. As a reasonable approximation, we can use the carrier density and density of states of the zero-field band structure~\cite{arai2009quantum,gobel2017QHE}; shown in Fig.~\ref{fig:m5bulklarge}(e). Note however that this simplified transport theory assumes large skyrmions for which the Landau levels are dense as for the case of $\lambda=10a$ presented in Fig.~\ref{fig:m5bulklarge}.}

\change{In the upper block of the band structure, where the conduction electrons feel a positive emergent field, the topological Hall conductivity is given by the carrier density $n(E)$. Note that we account for the electron-like and hole-like character of the states by the sign of $n(E)$. In the lower block, the topological Hall conductivity is reversed. In total, we find $\sigma_{xy,\mathrm{approx}}(E)\propto\pm n(E)$ for the two blocks, respectively. The spin Hall conductivity is roughly given by the product of the topological Hall conductivity and the out-of-plane spin polarization that is opposite for the two blocks due to the opposite emergent field of the skyrmion acting on the corresponding conduction electrons. Therefore, $\sigma_{xy,\mathrm{approx}}^{S_z}(E)\propto n(E)$ for both blocks. Likewise, the orbital Hall conductivity is given as the product of the topological Hall conductivity and the out-of-plane orbital angular momentum which scales with $\pm n/n'$. Therefore, $\sigma_{xy,\mathrm{approx}}^{L_z}(E)\propto [n(E)]^2/[n'(E)]$ for both blocks~\cite{gobel2024OHE}.} 

\change{All three approximate dependencies have been added as gray curves to Fig.~\ref{fig:m5bulklarge}(b-d). The approximation assumes a large smooth skyrmion texture with a homogeneous emergent field. The assumptions are not strictly fulfilled which shows in the plateaus in the Hall conductivities that are not resembled and the small deviations near the center of each block. Note also the following technical detail: We have modified the hopping amplitude for the calculation of the density of states in Fig.~\ref{fig:m5bulklarge}(e) to $\tilde{t}=\overline{\cos(\theta_{ij}/2)}\, t\approx 0.9775\, t$ to account for the finite size of the skyrmion, as in Ref.~\cite{hamamoto2015quantized}. $\theta_{ij}$ is the angle between two neighboring moments of the skyrmion.}

\change{Overall, the simplified explanation based on the density of states resembles the topological Hall, spin Hall and orbital Hall conductivities well. Also, it allows us to understand the scaling of the effects with the skyrmion area. The orbital Hall conductivity surpasses the spin Hall conductivity by orders of magnitude because $L_z$ can in principle have arbitrarily high values while $S_z$ is limited by $\pm\hbar/2$. $L_z$ scales roughly linearly with the skyrmion area because larger skyrmions allow for larger orbits. This leads to the quadratic dependence of the orbital Hall conductivity on the skyrmion area presented in Fig.~S2(b) in the Supplementary Material.} 

\change{To further confirm the validity of our calculations, we have simulated the effect of disorder by adding random onsite energies in the range of $[-0.2\,\mathrm{eV}, 0.2\,\mathrm{eV}]$. As a result, we see very similar conductivities, as presented in Fig.~S3 in the Supplementary Material.}

\textbf{Edge states.}
By considering a slab geometry, we gain information about edge states contributing to the different transverse transport phenomena. We repeat the skyrmion unit cell 10 times along the $y$ direction and keep periodicity along the $x$ direction. The corresponding band structure is shown in Fig.~\ref{fig:edge}a. Additionally, we have superimposed $L_{\nu,z}(k_\parallel)$ as a color code (red and blue) and the projected bulk bands in gray. Edge states are present that bridge the gaps between the bulk states. As the main result, they are orbital and spin polarized. 

\begin{figure*}[t!]
    \centering
    \includegraphics[width=\textwidth]{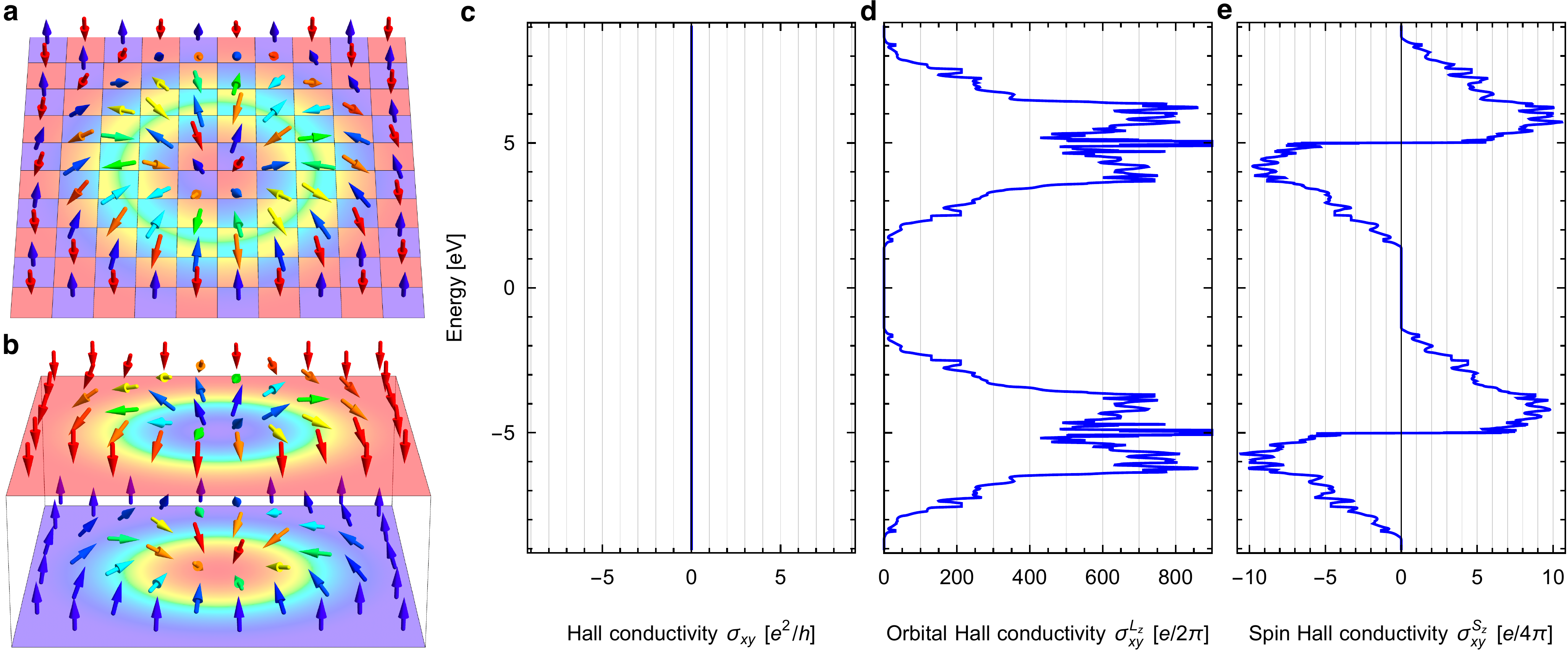}
    \caption{\textbf{Pure topological orbital Hall effect in antiferromagnetic skyrmion crystals.} \textbf{a} Antiferromagnetic skyrmion in one layer. The color resembles the out-of-plane orientation of the magnetic moments (arrows). \textbf{b} Synthetic antiferromagnetic skyrmion in two layers. \textbf{c} Hall conductivity $\sigma_{xy}$, \textbf{d} orbital Hall conductivity $\sigma_{xy}^{L_z}$ and \textbf{e} spin Hall conductivity $\sigma_{xy}^{S_z}$ as a function of energy in an antiferromagnetic skyrmion crystal characterized by second-nearest neighbor hopping $t_2=-1\,\mathrm{eV}$ and a Hunds's coupling strength of $m=5\,\mathrm{eV}$ and skyrmion size $\lambda=8a$.}
    \label{fig:afmskyrmion}
\end{figure*}

The discussion becomes easier if we are in the adiabatic limit where $m\gg |t|$. Therefore, Fig.~\ref{fig:edge}b shows the upper block for $m = 900|t| = 900\,\mathrm{eV}$. Similar to the Landau levels appearing in the bulk, the edge states are orbital polarized explaining the origin of the orbital Hall conductivity. In the gap between the lowest two bands in the upper block, we observe two states labeled (i) and (ii). Fig.~\ref{fig:edge}c shows that they are located at opposite edges of the slab. Their orbital angular momentum is positive and the spin and charge are negative. This corresponds to the positive orbital Hall conductivity and negative topological Hall and spin Hall conductivities calculated for the bulk at that energy. A schematic interpretation in terms of skipping orbits is shown in Fig.~\ref{fig:edge}d: By combining a translational and rotational degree of freedom, orbital angular momentum is transported along the edge as an orbital current.

\change{\textbf{Experimental detection.}}
\change{The detection of the orbital Hall conductivity is challenging because orbital currents cannot be measured directly. Three indirect approaches have been identified in the literature~\cite{jo2024spintronics, gobel2024OHE}: (i) The orbital currents are injected into an attached ferromagnet where the spin-orbit coupling transforms them into spin currents that exhibit a torque onto the magnetization that can be measured~\cite{lee2021orbital}. (ii) As our slab calculations reveal, edge states occur that are orbital polarized. These states, as well as accumulation of orbital angular momentum can be measured by the magneto-optical Kerr effect (MOKE)~\cite{choi2023observation,lyalin2023magneto}. (iii) The inverse orbital Hall effect can be measured by injecting orbital currents and measuring the charge current as a response, similar to how the inverse orbital Edelstein effect was measured~\cite{el2023observation}.}

\change{All three methods are applicable to observe the topological orbital Hall effect in skyrmion textures. However, besides the orbital contribution, a spin contribution emerges as well. Our calculations presented in Fig.~\ref{fig:m5bulk} have revealed that the orbital Hall conductivity is always much larger than the spin Hall conductivity in a skyrmion sample. Still, it is important to distinguish the individual contributions. For this reason, we can utilize the dependence of the Hall conductivities on the skyrmion size; cf. Fig.~S2 in the Supplementary Material. While the spin Hall conductivity scales linearly with the skyrmion area, just as the topological Hall conductivity, the orbital Hall conductivity scales quadratically with the skyrmion area. This is because the spin is limited by the quantum number $S=1/2$ but the orbital angular momentum is not and increases with the skyrmion area. Due to this characteristic size dependence, the orbital Hall conductivity can be extracted in an experiment by changing the skyrmion density via a change in temperature or the external magnetic field. As we will present in the following, this analysis becomes easier for alternative magnetic textures beyond skyrmions, as some of those textures do not exhibit a topological Hall effect or spin Hall effect.}

\textbf{Different types of skyrmions.}
\change{Different types of skyrmions have been observed in B20 materials such as MnSi~\cite{muhlbauer2009skyrmion}, magnetic multilayers like Ir/Fe/Co/Pt~\cite{soumyanarayanan2017tunable}, centrosymmetric materials like Sc-doped barium
ferrite~\cite{yu2012magnetic}, insulating multiferroics like Cu$_2$OSeO$_3$~\cite{seki2012observation} and others.} The above discussed calculations can be repeated for other non-collinear spin textures related to skyrmions~\cite{gobel2021beyond}. We start by discussing other objects with a finite topological charge. To classify such objects, one can introduce the topological charge $N_\mathrm{Sk}=pv$ that is the product of polarity $p=\pm 1$ (out-of-plane orientation of the center magnetic moment) and vorticity $v=0,\pm 1, \pm 2, ...$. The latter quantity relates the polar angle of the position vector $\varphi$ with the polar angle of the magnetic texture $\phi$ via $\phi=v\varphi+\gamma$.
Here, $\gamma$ is an offset that is called helicity. Since the topological charge is independent of the helicity, Néel ($\gamma=0,\pi$) and Bloch skyrmions ($\gamma=\pm\pi/2$) of otherwise equal profile and electronic properties exhibit the same topological Hall, spin Hall and orbital Hall responses.

So far, we have discussed skyrmions that are characterized by $p=1$ and $v=1$ so that they carry a topological charge of $N_\mathrm{Sk}=+1$. As a consequence, their emergent field for parallel spin-alignment points along $+z$. If we consider the same skyrmions in a ferromagnetic background that is oriented oppositely, the polarity changes sign $p=-1$ and so does the topological charge $N_\mathrm{Sk}=-1$. As a consequence, the emergent field points along $-z$ and so the topological Hall conductivity changes sign. However, the spin Hall conductivity and the orbital Hall conductivity remain invariant because $S_z$ and $L_z$ change sign as well, due to the reversed skyrmion texture and emergent field, respectively.

A similar discussion holds for antiskyrmions \change{that have been observed in the Heusler material MnPtSn}~\cite{nayak2017magnetic} that are characterized by a negative vorticity $v=-1$. Antiskyrmions and skyrmions in the same magnetic background (same polarity $p$) have opposite topological charges and therefore opposite emergent fields. As a consequence, they exhibit opposite topological Hall \change{and spin Hall} conductivities but the same orbital Hall conductivities. If skyrmions and antiskyrmions coexist at equal numbers, as is possible in frustrated magnets~\cite{okubo2012multiple} or Heusler materials~\cite{peng2020controlled,jena2020elliptical,jena2020evolution,gobel2021quaternary}, the net topological Hall \change{and spin Hall conductivities are} compensated but the orbital Hall effect is not which makes it ideal to detect topological magnetic textures if one is not sure about the type of texture.

\change{Fig.~S4 in the Supplementary Material shows the numerically calculated Hall conductivities of a crystal of Bloch skyrmions, antiskyrmions and bimerons (in-plane skyrmions)~\cite{kharkov2017bound, gobel2019magnetic}. All three textures have been constructed such that they have the same (or opposite) topological charge density as the Néel skyrmion presented in Fig.~\ref{fig:m5bulk}. As long as no spin-orbit coupling is considered, the Néel skyrmion, Bloch skyrmion and bimeron have the same topological Hall conductivity and the antiskyrmion has the exactly opposite topological Hall conductivity due to the opposite topological charge density. The calculated spin Hall conductivitiy is equal for the Néel and Bloch skyrmions and again opposite for the antiskyrmion. The bimeron exhibits a vanishing spin Hall conductivity $\sigma_{xy}^{S_z}$ because the out-of-plane net moment of the bimeron is zero. Most importantly, all 4 textures exhibit exactly the same orbital Hall conductivity.}  

The idea of coupling two skyrmionic textures with opposite topological charges gave rise to other textures such as the skyrmionium~\cite{zhang2016control,zhang2018real,gobel2019electrical}, the antiferromagnetic skyrmion~\cite{barker2016static,zhang2016magnetic,zhang2016antiferromagnetic,gobel2017afmskx,legrand2020room,dohi2019formation} \change{or the antiferromagnetic bimeron~\cite{gobel2019magnetic, gobel2021beyond, shen2020current}}. In the following, we will discuss the case of the antiferromagnetic skyrmion in detail.


\textbf{Pure topological orbital Hall effect in antiferromagnetic skyrmions.}
To model an antiferromagnetic skyrmion, we take the texture considered before and reverse every second magnetic moment in a checkerboard pattern (cf. Fig.~\ref{fig:afmskyrmion}a). This allows to distinguish two skyrmions on two sublattices with opposite magnetic moments and topological charges leading to a compensated topological charge overall. 

Besides this true antiferromagnetic skyrmion~\cite{barker2016static,zhang2016antiferromagnetic,gobel2017afmskx}, synthetic antiferromagnetic skyrmions (cf. Fig.~\ref{fig:afmskyrmion}b) have been considered as well~\cite{zhang2016magnetic} \change{and have even been observed experimentally in multilayer systems~\cite{dohi2019formation,legrand2020room}.} In the latter case, the interpretation of two coexisting sub-skyrmions is even more applicable. This is especially true once we extend the $s$-$d$ Hamiltonian to first\change{-nearest neighbor hopping with amplitude $t_1$ (the hopping between the different sublattices)} and second-nearest neighbor hopping \change{with amplitude $t_2$ (the hopping within the sublattices)} and consider only second-nearest neighbor hopping $t_2$. 

The corresponding charge, orbital and spin Hall conductivites for $t_1=0$ and $t_2=-1\,\mathrm{eV}$ are shown in Fig.~\ref{fig:afmskyrmion}c-e. The bands are doubly degenerate resembling the opposite spin alignment in the two sublattices hosting skyrmions with opposite emergent magnetic fields. As a consequence, the topological Hall effect is compensated for every energy (Fig.~\ref{fig:afmskyrmion}c) but the orbital and spin Hall conductivities are finite (Fig.~\ref{fig:afmskyrmion}d,e). Their energy dependencies are similar to those of the conventional skyrmion presented in Fig.~\ref{fig:m5bulk}c,d but their magnitude is larger due to the two coexisting sublattices and the larger skyrmion size. Qualitatively similar results can be obtained for nearest-neighbor hopping only ($t_1=-1\,\mathrm{eV}$ and $t_2=0$, presented in Fig.~S5a-c of the Supplementary Material) and both hoppings considered ($t_1=-2/3\,\mathrm{eV}$ and $t_2=-2/3\,\mathrm{eV}$, presented in Fig.~S5d-f of the Supplementary Material). 

Independent of the choice of hopping amplitudes, a pure Hall effect of angular momentum (orbital and spin) emerges while the charge transport is compensated. The precise energy dependence, however, can be very different once the hoppings are modified because the electronic structure changes.

\change{Fig.~S6 in the Supplementary Material shows equivalent results for an antiferromagnetic bimeron~\cite{gobel2019magnetic, gobel2021beyond, shen2020current}. Like for the antiferromagnetic skyrmion discussed above, the topological Hall effect vanishes due to a compensated topological charge. However, since the two bimerons do not have a net out-of-plane moment, the spin Hall effect is zero as well. Therefore, the antiferromagnetic bimeron, that has been observed experimentally~\cite{jani2021antiferromagnetic,bhukta2024homochiral}, might be the ideal platform to investigate the topological orbital Hall effect, as it exhibits a pure orbital Hall effect that is not superimposed by a charge or spin Hall effect.}\\
%
%
%
%
%
\\
\noindent\textbf{Conclusions}\\
In summary, we have shown that skyrmions and antiferromagnetic skyrmions give rise to a topological orbital Hall effect. For conventional skyrmions, the net emergent field forces electrons onto circular trajectories and causes skipping orbits at the edges akin to the quantum Hall effect. This gives rise to a topological Hall effect for which the charge currents are spin and orbital polarized. An antiferromagnetic skyrmions on the other hand, consists of two oppositely oriented subskyrmions on two sublattices. The opposite emergent fields cause opposite topological Hall effects and opposite spin and orbital polarization giving rise to a compensated topological Hall effect but net spin and orbital Hall effects. \change{An antiferromagnetic bimeron even exhibits a pure orbital Hall effect.}

Since the orbital angular momentum \change{is not restricted by a quantum number but increases with the skyrmion area}, the orbital Hall conductivity is much larger than the spin Hall conductivity. The orbital Hall conductivity scales roughly quadratically with the skyrmion \change{area} while the spin and charge Hall conductivities scale roughly linearly \change{(cf. Fig.~S2 of the Supplementary Material)}.
This means these textures could serve as generators of large orbital currents that can potentially transport information and give rise to considerable orbital torques~\cite{go2020orbital,lee2021orbital}.

The results presented for the skyrmion can be carried over to other topological spin textures such as antiskyrmions~\cite{nayak2017magnetic}, bimerons~\cite{kharkov2017bound,gobel2019magnetic,gao2019creation} and biskyrmions~\cite{yu2014biskyrmion,gobel2019forming}. Likewise, the results of the antiferromagnetic skyrmion can be carried over to other compensated topological spin textures such as skyrmioniums~\cite{zhang2016control,zhang2018real,gobel2019electrical}.

\change{It is worth noting again that in the present study we have focused on the intrinsic contribution of a skyrmionic texture to the orbital Hall effect in an easy-to-understand model system. We have restricted the model to $s$ orbitals, have disregarded spin-orbit coupling and did not consider scattering at defects. Taking these effects into account in a future study could be interesting as additional effect like the anomalous Hall effect or extrinsic effects like the `side jump' and `skew scattering' might occur that can be important corrections~\cite{liu2024quantum, liu2024dominance}. Furthermore, it has been shown recently that the anomalous position has to be considered in the definition of the velocity operator once the system includes more than just $s$ orbitals~\cite{go2024first}. As explained in Ref.~\cite{atencia2024orbital}, a complete quantum mechanical theory of non-equilibrium orbital angular momentum dynamics is not yet available and has to be derived from the ground up in the future.}\\
%
%
%
%
\\
\textbf{Methods}\\
\textbf{Skyrmion texture.}
A skyrmion centered at $\vec{r}_0=0$ is modelled by 
\begin{align}
    \vec{m}=\begin{pmatrix}
        x\sin(2\pi r/ \lambda )/r\\
        y\sin(2\pi r/ \lambda )/r\\
        \cos(2\pi r/ \lambda )
    \end{pmatrix}
\end{align}
for $r=\sqrt{x^2+y^2}<\lambda/2$ and $\vec{m}=-\vec{e}_z$ otherwise.

Note that this is only an approximation for the normalized skyrmion profile. The precise magnetization profile is determined by the interplay of several magnetic interactions and depends on the magnetic parameters. Most important for our purposes is the topological charge~\cite{nagaosa2013topological}
\begin{align}
	N_\mathrm{Sk}  =  \frac{1}{4\pi} \int_{xy} \vec{m}(\vec{r}) \cdot \left[ \frac{\partial \vec{m}(\vec{r})}{\partial x}  \times  \frac{\partial \vec{m}(\vec{r})}{\partial y}  \right]\, \mathrm{d}^{2} r.
\end{align}
which is an integer for the above considered Néel skyrmion, $N_\mathrm{Sk} = +1$.\\
\\
\textbf{Calculation of the observables.}
The spin is calculated as 
\begin{align}
    S_{z,\nu}(\vec{k})=\frac{\hbar}{2}\braopket{\nu\vec{k}}{\sigma_z}{\nu\vec{k}}.
\end{align}
We calculate the orbital angular momentum based on the modern formulation including the off-diagonal elements of the tensor~\cite{pezo2022orbital,gobel2024OHE}
\begin{align}
      &\braopket{\nu \vec{k}}{L_z}{\alpha \vec{k}} = \mathrm{i} \frac{e\hbar^2}{4g_L\mu_\mathrm{B}}  \sum_{\beta \neq \nu, \alpha} \left( \frac{1}{E_{\beta \vec{k}} - E_{\nu \vec{k}}} + \frac{1}{E_{\beta \vec{k}} - E_{\alpha \vec{k}}} \right)\notag\\
      &\times\left(\braopket{\nu \vec{k}}{v_x}{\beta \vec{k}} \braopket{\beta \vec{k}}{v_y}{\alpha \vec{k}} - \braopket{\nu \vec{k}}{v_y}{\beta \vec{k}} \braopket{\beta \vec{k}}{v_x}{\alpha \vec{k}}\right). \label{EQ:Lz_matrix_elements}
\end{align}
Here, $\vec{v}=\frac{1}{\hbar}\nabla_{\vec{k}}H$ is the velocity operator. $L_{\nu,z}(\vec{k})$ are the diagonal elements of the tensor $L_{\nu,z}(\vec{k})=\braopket{\nu \vec{k}}{L_z}{\nu \vec{k}}$. \change{Note that we have corrected a mistake in Ref.~\cite{pezo2022orbital} where $\mathrm{Im}$ was used instead of the imaginary unit $\mathrm{i}$.}

The intrinsic Hall conductivity~\cite{nagaosa2010anomalous}
\begin{align}
    \sigma_{xy}(E_\text{F})= -\frac{e^2}{h}\sum_\nu \frac{1}{2\pi}\int_{E_{\nu \vec{k}}\leq E_\text{F}}\Omega_{\nu,z}(\vec{k}) \,\mathrm{d}^2k\label{EQ:sigma_xy_Kubo}
\end{align}
is calculated by integrating the reciprocal space Berry curvature over all occupied states in the Brillouin zone (states below the Fermi energy $E_\mathrm{F}$ at zero temperature). The Berry curvature is~\cite{berry1984quantal}
\begin{align}
    \Omega_{\nu,z}(\vec{k})= -2 \hbar^2\ \text{Im}\ \sum_{\mu\neq \nu} \frac{\braopket{\nu \vec{k}}{v_x}{\mu \vec{k}} \braopket{\mu \vec{k}}{v_y}{\nu \vec{k}}}{(E_{\nu \vec{k}} - E_{\mu \vec{k}})^2}.
\end{align}
The intrinsic orbital and spin Hall conductivities~\cite{pezo2022orbital}
\begin{align}
    \sigma^{L_z}_{xy}(E_\text{F})&= \frac{e}{\hbar}\sum_\nu \frac{1}{(2\pi)^2}\int_{E_{\nu \vec{k}}\leq E_\text{F}}\Omega_{\nu,z}^{L_z}(\vec{k}) \,\mathrm{d}^2k,\label{EQ:sigma_Lz_xy_Kubo}\\
    \sigma^{S_z}_{xy}(E_\text{F})&= \frac{e}{\hbar}\sum_\nu \frac{1}{(2\pi)^2}\int_{E_{\nu \vec{k}}\leq E_\text{F}}\Omega_{\nu,z}^{S_z}(\vec{k}) \,\mathrm{d}^2k\label{EQ:sigma_Sz_xy_Kubo}
\end{align}
are calculated from the orbital and spin Berry curvatures, respectively, 
\begin{align}
    \Omega_{\nu,z}^{L_z}(\vec{k})&= -2 \hbar^2\ \text{Im}\ \sum_{\mu\neq \nu} \frac{\braopket{\nu \vec{k}}{j_x^{L_z}}{\mu \vec{k}} \braopket{\mu \vec{k}}{v_y}{\nu \vec{k}}}{(E_{\nu \vec{k}} - E_{\mu \vec{k}})^2},\\
    \Omega_{\nu,z}^{S_z}(\vec{k})&= -2 \hbar^2\ \text{Im}\ \sum_{\mu\neq \nu} \frac{\braopket{\nu \vec{k}}{j_x^{S_z}}{\mu \vec{k}} \braopket{\mu \vec{k}}{v_y}{\nu \vec{k}}}{(E_{\nu \vec{k}} - E_{\mu \vec{k}})^2},
\end{align}
where $\braopket{\nu \vec{k}}{j_x^{L_z}}{\mu \vec{k}} = \frac{1}{2}\sum_{\alpha} [ \braopket{\nu \vec{k}}{v_x}{\alpha \vec{k}} \braopket{\alpha \vec{k}}{L_z}{\mu \vec{k}} + \braopket{\nu \vec{k}}{L_z}{\alpha \vec{k}} \braopket{\alpha \vec{k}}{v_x}{\mu \vec{k}} ]$ is the orbital current operator and $\braopket{\nu \vec{k}}{j_x^{S_z}}{\mu \vec{k}} = \frac{1}{2} [ \braopket{\nu \vec{k}}{v_x}{\nu \vec{k}} \braopket{\mu \vec{k}}{S_z}{\mu \vec{k}} + \braopket{\nu \vec{k}}{S_z}{\nu \vec{k}} \braopket{\mu \vec{k}}{v_x}{\mu \vec{k}} ]$ the spin current operator.\\
\\
\textbf{Data availability}\\
The data that support the findings of this work are available at \changeb{DOI: 10.5281/zenodo.13919926.}\\
\\
\textbf{Code availability}\\
The code that supports the findings of this work is available from the authors on reasonable request.\\
\\
\textbf{Acknowledgements}\\
This work was supported by the EIC Pathfinder OPEN grant 101129641 ``Orbital Engineering for Innovative Electronics''.\\
\\
\textbf{Author contributions}\\
B.G. and L.S. performed calculations.
B.G. wrote the manuscript with significant inputs from all authors.
B.G. prepared the figures.
All authors discussed the results.
B.G. planned the project.
B.G. and I.M. supervised the project.
I.M. provided the funding.\\
\\
\textbf{Supplementary information}\\
accompanies this paper at [insert link].\\
\\
\textbf{Competing interests}\\
The authors declare no competing interests.\\
\\
\changeb{\textbf{References}}




\end{document}